\documentclass{article}
\usepackage{ijcai23}

\usepackage{times}
\usepackage{soul}
\usepackage{url}
\usepackage[hidelinks]{hyperref}
\usepackage[utf8]{inputenc}
\usepackage[small]{caption}
\usepackage{graphicx}
\usepackage{amsmath}
\usepackage{amsthm}
\usepackage{booktabs}
\usepackage{algorithm}
\usepackage{algorithmic}
\usepackage{bm}
\usepackage{amssymb}
\usepackage{threeparttable}
\usepackage{booktabs}
\usepackage{multirow}
\usepackage{caption}
\usepackage{subcaption}
\usepackage{bbding}
\usepackage[T1]{fontenc}

\title{Scalable Communication for Multi-Agent Reinforcement Learning \\
via Transformer-Based Email Mechanism}

\author{
    Xudong Guo\and Daming Shi\and Wenhui Fan
    \affiliations
    Department of Automation, Tsinghua University
    \emails
    \{gxd20, shidm18\}@mails.tsinghua.edu.cn, fanwenhui@tsinghua.edu.cn
}
\begin{document}

\maketitle

\begin{abstract}
    Communication can impressively improve cooperation in multi-agent reinforcement learning (MARL), especially for partially-observed tasks. However, existing works either broadcast the messages leading to information redundancy, or learn targeted communication by modeling all the other agents as targets, which is not scalable when the number of agents varies. In this work, to tackle the scalability problem of MARL communication for partially-observed tasks, we propose a novel framework \textbf{Transformer-based Email Mechanism (TEM)}. The agents adopt local communication to send messages only to the ones that can be observed without modeling all the agents. Inspired by human cooperation with email forwarding, we design message chains to forward information to cooperate with the agents outside the observation range. We introduce Transformer to encode and decode the message chain to choose the next receiver selectively. Empirically, TEM outperforms the baselines on multiple cooperative MARL benchmarks. When the number of agents varies, TEM maintains superior performance without further training.
\end{abstract}

\section{Introduction}
Multi-agent reinforcement learning (MARL) has achieved remarkable success in many complex challenges, especially in game playing \cite{openai_dota_2019,vinyals_grandmaster_2019}. MARL shows great potential to solve cooperative multi-agent real-world tasks, such as autonomous vehicle teams \cite{shalev-shwartz_safe_2016},  robotics control \cite{kober_reinforcement_2013} and intelligent
traffic control \cite{wei_colight_2019}. However, some essential obstacles still exist for MARL to reach satisfactory performance. When training the MARL algorithms, the agents keep updating their policies and causing dynamics in the environment, which may hinder the model convergence. Worse still, in most cooperative multi-agent tasks, agents can only observe part of the environment. Partial observability and non-stationarity make it harder to successfully cooperate, even though some works employ centralized training and decentralized execution (CTDE) paradigm to import a critic to coordinate the whole team \cite{yu_surprising_2021,lowe_multi-agent_2020,son_qtran_2019,rashid_qmix_2018}. 
\begin{figure}[t]
% \vspace{-0.5cm}
\centering
\includegraphics[width=0.5\columnwidth]{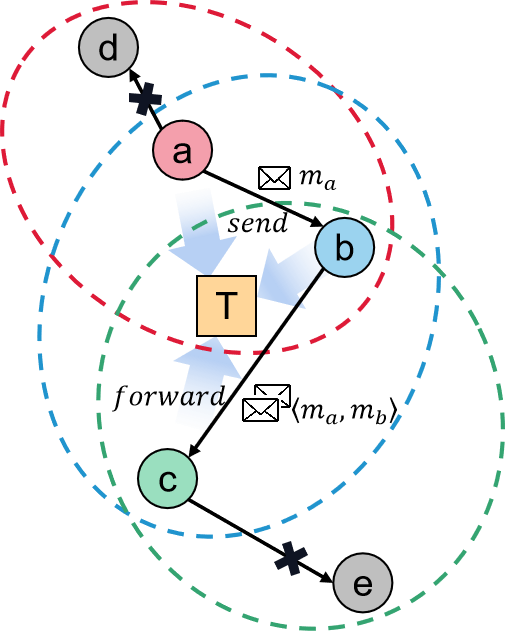} 
\caption{Message chain formed by email forwarding in the Transformer-based Email Mechanism (TEM). The agents (circles) are trying to surround and capture the target (square). The dotted circle is the observation range for the agent with the same color. The black lines are message chains. $\langle \cdot \rangle$ denotes concatenating. The cross denotes not sending or forwarding after receiver selection. The agents \textit{a}, \textit{b} and \textit{c} indirectly cooperate by sending ($m_a$) and forwarding ($\langle m_a,m_b \rangle$) messages to capture the target \textit{T}. As they are not in the same observation range, forwarding like emails is necessary.}
\label{fig0}
\vspace{-0.5cm}
\end{figure}
Inspired by the ways how humans and animals cooperate, communication is introduced to share information between agents. Some works broadcast the messages to all the other agents \cite{zhang_efficient_2019,sukhbaatar_learning_2016,foerster_learning_2016}, and other recent works try to learn targeted peer-to-peer communication to reduce the communication bandwidth \cite{ding_learning_2020,jiang_learning_2018,yuan_multi-agent_2022}. Attention mechanism from Transformer \cite{vaswani_attention_2017} is also employed to learn the communication \cite{jiang_learning_2018}. However, the existing methods rely on modeling every teammate in the environment by ID to decide whether to communicate, which will bring huge computational overhead when the number of agents is large. As the modeling network is trained by a specific amount of IDs, the learned communication mechanism is not scalable to reuse when the number of agents changes. In fact, the agent cannot know the state of an agent outside the observation range, and cannot judge whether the information is useful for it, so it is unreasonable to directly share information with such an agent. 
For example, 
for applications to autonomous vehicles, only the vehicles nearby are worth communicating with to avoid collisions. 
% Thus, global communication with all vehicles is unnecessary.
Moreover, communication with other vehicles should adapt to different numbers of agents as the traffic situation varies a lot.

In this work, to tackle this \textbf{new problem} - the scalability of MARL communication, we propose a scalable multi-agent communication mechanism via Transformer-based Email Mechanism (TEM) to tackle the abovementioned challenges as shown in Fig \ref{fig0}. We adopt local communication to send messages only to the agents in the observation range, without modeling all the agents.
% As the agents in the range may change, we do not use the absolute ID among all the agents in the environment to represent the communication object. Instead, we use local communication among the agents in the range. 
The agent will decide whom to communicate with by its own intention and by observing the agents in the range. Thus, no matter how the overall number of the agents changes, the learned communication mechanism is scalable. To better utilize the key information and indirectly cooperate with the agents outside the range, we design a \textbf{new communication mechanism} like email forwarding to form a message chain. The agents can send and forward the messages so that the chain connects agents from different ranges. For example, the agent \textit{a} in Fig \ref{fig0} would like to surround and capture the target \textit{T} with other agents, thus the agent \textit{a} may send a message to the agent \textit{b} instead of \textit{d}, though \textit{d} is the nearest. Then the agent \textit{b} can forward the message together with the information from itself to \textit{c}, so that \textit{a}, \textit{b} and \textit{c} can cooperate for the same goal though there is no direct communication between them. Similarly, in our daily life, cooperation in a big company or organization relies on such forwarding emails to share information, as it is always hard to directly find the exact contact in another department.

To suit the unfixed length of the message chain and ensure the communication mechanism is scalable, we design a \textbf{new message network} and employ Transformer to encode and decode the sequence of messages. Furthermore, augmented by the attention mechanism in the Transformer, the communication is selective by modeling the correlation between the messages and the observation. The message network is independent and can be plugged into any CTDE method. What's more, we design a loss to guide the agent to estimate the impact of the message on other agents. Note that we do not introduce the broadcast mechanism from email to keep the communication efficient. 

For evaluation, we test TEM on three partial-observation cooperative MARL benchmarks: the Starcraft Multi-Agent Challenge (SMAC) \cite{samvelyan_starcraft_2019}, Predator Prey (PP) \cite{kim_learning_2019} and Cooperative Navigation (CN) \cite{lowe_multi-agent_2020}, where TEM reaches better performance than the baselines. We also evaluate the scalability of TEM. Without extra training, TEM can suit both situations where the number of agents increases and decreases, and still outperforms the baselines. 

In sum, the proposed TEM has three advantages: \textcircled{1} \textbf{decentralized} without a central communication scheduler, which is more practical; \textcircled{2} \textbf{selective} to send and forward information to the agent who needs it most, while other methods use broadcast which may bring redundant information and the communication cost is high; \textcircled{3} \textbf{scalable} without retraining since TEM does not model all the agents as other methods.

\section{Related Works}
Learning how to communicate is a popular research domain in multi-agent reinforcement learning. Research in this domain focus mainly on cooperative scenarios, where agents could communicate with each other explicitly. In the early works, RIAL and DIAL \cite{foerster_learning_2016} are designed to learn communication, where messages are delivered from one timestep to the next timestep in a broadcast way. CommNet \cite{sukhbaatar_learning_2016} proposes a hidden layer as communication and allows the agents to communicate repeatedly in each step. IC3Net \cite{singh_learning_2018} brings in the gating mechanism to control communication based on CommNet. Both of BiC-Net \cite{peng_multiagent_2017} and ATOC \cite{jiang_learning_2018} implement the communication layer as bidirectional RNN, which inputs the observations of all agents and outputs the action or the integrated thought of each agent. However, these methods either broadcast the messages or rely on a centralized communication layer, which is high-cost and not stable. Communication should not only serve as an integration of information, 
instead, the agents should share information selectively through peer-to-peer communication.

To avoid broadcasting messages, recent works try to design more intelligent communication mechanisms. CTDE paradigms are also imported to implement decentralized communication. Some works are based on QMIX \cite{rashid_qmix_2018}: VBC \cite{zhang_efficient_2019} proposes a request-reply mechanism and a variance-based regularizer to eliminate the noisy components in messages.
% NDQ \cite{wang_learning_2020} learns nearly decomposable value functions with communication.
TMC \cite{zhang_succinct_2020} maximizes the mutual information between the decentralized Q functions and the communication messages while minimizing the entropy of messages between agents. MAIC \cite{yuan_multi-agent_2022} allows each agent to learn to generate incentive messages by modeling every teammate and bias other agents' value functions directly. Some are based on another CTDE framework MADDPG \cite{lowe_multi-agent_2020}: TarMAC \cite{das_tarmac_2019} proposes a targeted communication behavior via a signature-based soft attention mechanism. Besides the message, the sender broadcasts a key used by the receivers to gauge the message's relevance.
% DAACMP \cite{mao_learning_2020} adds a double attention mechanism in the actor and critic network respectively to select and process the important messages.
I2C \cite{ding_learning_2020} learns a prior net via causal inference for peer-to-peer communication. The influence of one agent on another is inferred via the joint action-value function and quantified to label the necessity of peer-to-peer communication. Nevertheless, the methods above need to model every other agent in the environment to achieve individual communication, which is not scalable and practical.

To the best of our knowledge, none of the existing MARL communication methods considers the communication scalability and the forwarding protocol inspired by email.

\section{Background}
% \subsection{Problem Formation}
\subsection{Policy Gradient (PG) Reinforcement Learning}
Policy Gradient (PG) reinforcement learning has the advantage of learning a policy network explicitly, in contrast to value-based reinforcement learning methods. PG methods optimize the policy parameter $\theta$ to maximize its objective function $J(\theta)=E_S(V_{\pi_\theta}(s))$.
% , where the $\theta$ denotes the parameters of policy function. 
% Give the state $s_t$, we have the gradient of objective function $g(\theta)=\frac{\partial V_\pi(s_t)}{\partial\theta}=\frac{\partial E_A[\pi_\theta(A|s_t)Q_\pi(s_t,A)]}{\partial \theta}=E[\frac{\partial log_\pi_\theta(A|s_t)}{\partial \theta}Q_\pi(s_t,A)]=E[\frac{\partial log_\pi_\theta(A|s_t)}{\partial \theta}(Q_\pi(s_t,A)-V_\pi(s_t))]$. Therefore, if an action $a$ is taken based on policy $\pi_\theta$, the random gradient is $g(\theta)=\frac{\partial log_\pi_\theta(a|s_t)}{\partial \theta}(Q_\pi(s_t,a)-V_\pi(s_t))$. Then, given a learning rate $r$, the iterative policy parameters are $\theta\leftarrow \theta + rg(\theta)$. 
However, due to the variance of environments, it is hard to choose a subtle learning rate in reinforcement learning. 
% Specifically, once the learning rate is too small, the optimization speed cannot be guaranteed. Whereas, a large learning rate will lead the policy optimization to value collapse, where the updated parameters rush over the current region in policy space.
To resolve this problem and ensure the safe optimization of policy learning, the Trust Region Policy Optimization (TRPO) \cite{schulman_trust_2015} increases the constraint of the parameter difference between policy updates. 
% Such constraint ensures the parameter changes in a small range, so that the collapse of value can be avoided and the policy can learn monotonically.
% Take $\pi_\theta$ as the policy function with parameter $\theta$, 
% The parameter update of TRPO is $\mathop\theta_{k+1}={\arg\max}_{\theta} L(\theta_k,\theta)~s.t. \bar{D}_{KL}(\theta||\theta_k)\leq\delta$, where $L(\theta_k,\theta) = E[\frac{\pi_\theta(a|s)}{\pi_{\theta_k}(a|s)}A_{\pi_{\theta_k}}(s,a)]$ is the approximation of the original policy gradient object $J(\theta)$ within the constraint of KL divergence.
% To verify the equivalence of such optimization object, we have:
% \begin{equation}
% \begin{aligned}
% % \begin{align}
% J(\theta)
%         &=E_s[V_\pi(s)]\\
%         &=E_s[E_a[Q_\pi(s,a)]]\\
%         &=E_s[\sum_{a\sim\theta_k}\pi_\theta(a|s)Q_{\pi_{\theta_k}}(s,a)]\\
%         &=E_s[\sum_{a\sim\theta_k}\pi_{\theta_k}(a|s) \frac {\pi_\theta(a|s)}{\pi_{\theta_k}(a|s)} Q_{\pi_{\theta_k}}]\\
%         &=E_{s,a}[\frac {\pi_\theta(a|s)}{\pi_{\theta_k}(a|s)} Q_{\pi_{\theta_k}}(s,a)],\\
% % \end{align}
% \end{aligned}
% \end{equation}
% given advantage function $A(s,a)=Q(s,a)-V(s)$\\
% \begin{equation}
% \begin{aligned}
%         \Rightarrow E_{s,a}[\frac {\pi_\theta(a|s)}{\pi_{\theta_k}(a|s)} A_{\pi_{\theta_k}}(s,a)]=L(\theta_k,\theta)
% \end{aligned}
% \end{equation}
% Although TRPO validates theoretically, the practical calculation is prohibitive for policy learning. Therefore, the simplified version 

Based on TRPO, a simplified version Proximal Policy  Optimization \cite{schulman_proximal_2017} is carried out, maintaining the motivation to constrain the learning step while more efficient and easy to be implemented. 
% In practice, there are two variants of PPO: PPO-penalty and PPO-clip.
% //
% \begin{itemize}
% 		\item PPO-penalty takes the KL divergence as a soft constraint instead of hard constrain. A regular term of KL divergence is added into the object function.
% 		\item PPO-clip drops this constraint, but keeps the update of $\theta$ in the learning step range with truncation.
% \end{itemize}
% In practice, the PPO-clip is more efficient and easy to implement. Therefore, our following work is all based on PPO-clip. 
The object function of PPO can be written as:
\begin{equation}
\begin{aligned}
\label{La}
\mathcal{L}(s,a,\theta_k,\theta)=&min[\frac {\pi_\theta(a|s)}{\pi_{\theta_k}(a|s)} A_{\pi_{\theta_k}}(s,a), \\
&clip(\frac {\pi_\theta(a|s)}{\pi_{\theta_k}(a|s)}),1-\epsilon,1+\epsilon)A_{\pi_{\theta_k}}(s,a)],
\end{aligned}
\end{equation}
which forces the ratio of $\frac {\pi_\theta(a|s)}{\pi_{\theta_k}(a|s)}$ to locate in the interval $(1-\epsilon,1+\epsilon)$, so that the new  $\theta$ is not too far away from old $\theta_k$.
% \subsection{Multi-agent Deep Reinforcement Learning}
\subsection{MAPPO Algorithm}
Multi-agent PPO (MAPPO) introduces PPO into the multi-agent scenario \cite{yu_surprising_2021}. MAPPO mainly considers decentralized partially observable Markov decision processes (DEC-POMDP). In an environment with $n$ agents, $s \in S$ denotes the state of the environment. The agent \textit{i} only has a local observation of environment $o_i=O(s)$ and chooses its action based on its observation and policy $a_i=\pi_i(a_i|o_i)$. The joint action $A=(a_1,...,a_n)$ denotes the set of actions of all agents. Then, the environment transits its state based on the transition probability $P(s' |s, A)$. In MARL, all the agents will get rewards based on the transition of state and their actions (or more likely joint action) $r_i=\mathcal{R}(s, A)$. Each agent is supposed to get a higher accumulated reward $\sum_t{r_i^t}$. Therefore, the agents optimize their policy to maximize the discount accumulated reward $J(\theta)=E_{a^t,s^t}[\sum_t {\gamma^{t} \mathcal{R}(s^t,a^t)}]$, where $\gamma \in(0,1]$ is the discount factor.

MAPPO utilizes parameter sharing within homogeneous agents, i.e., homogeneous agents share the same set of network structure and parameters during training and testing. MAPPO is also a CTDE framework, namely, each PPO agent maintains an actor network $\pi_\theta$ to learn the policy and a critic network $V_\phi(s)$ to learn the value function, where $\theta$ and $\phi$ are the parameters of policy network and value network, respectively.
% The policy network $\pi_\theta$ takes its action based on its observation and the value function $V_\phi(s)$ estimates the value of current state. Apparently,
The value function requires the global state and only works during training procedures to reduce variance. In our work, we take MAPPO as our baseline and backbone, and add TEM as the communication mechanism into MAPPO. 
% During the training phase, the critic network in CTDE will evaluate the learned action, while during the executing phase, the actor will make decision independently.
% It is noteworthy that TEM is agnostic to specific CTDE framework and can work on others like MADDPG. 
\section{Methods}
\begin{figure}[!htbp]
% \vspace{-0.5cm}
\centering
\includegraphics[width=0.9\columnwidth]{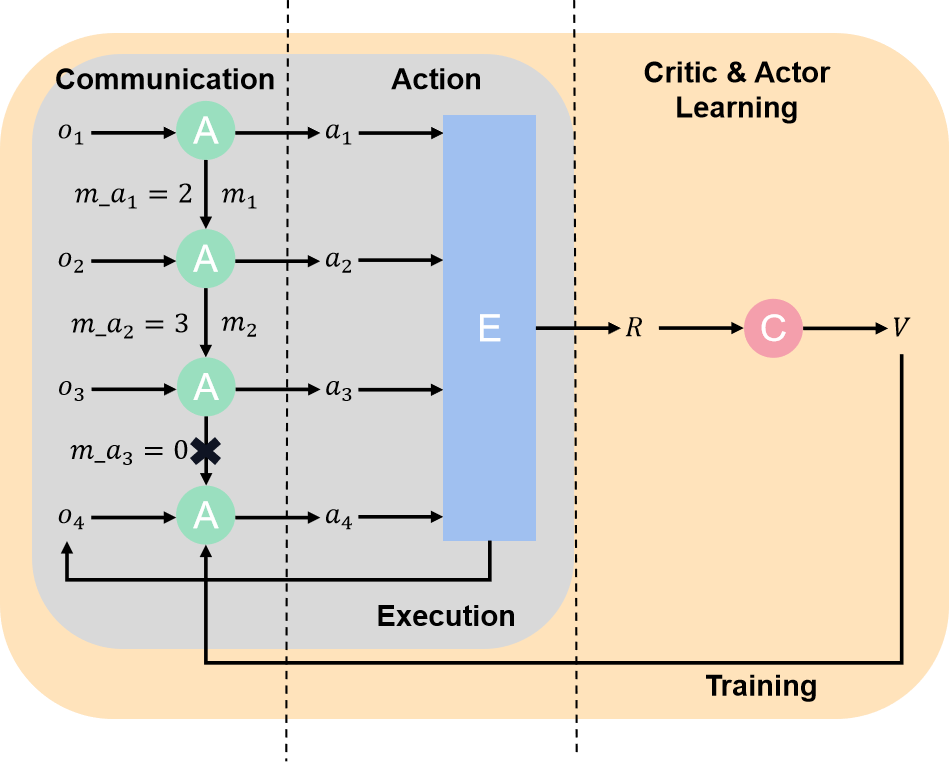} 
\caption{Workflow of TEM during one time step. A denotes the actor network, C denotes the critic network, E denotes the environment. One training step has three phases: communication, action and learning. The execution only includes the first two phases and the critic will not work.}
\label{fig1}
% \vspace{-0.5cm}
\end{figure}
In this section, we introduce the detailed structure and design of TEM. Before each action decision-making, the agents communicate with each other following the designed protocol, sharing the key information efficiently and selectively. We design a message module based on Transformer to encode the messages received. At the same time, the module is able to decide whether to communicate and whom to communicate with. The message module works together with the original action decision module from MAPPO, to form the actor network in the CTDE structure. The workflow of TEM is presented in Fig \ref{fig1} and Algorithm \ref{alg:algorithm}. We design an independent loss to encourage the message module to maximize the messages' impact on other agents. The whole model has the scalability to transfer from one scenario to another. As the message module is parallel to the action module, our model can be plugged into any CTDE structure.
\subsection{Communication Protocol Design}
We design a communication protocol following the way how humans communicate by email. The information flow is like a forwarding chain: the chain starts with an agent with key information to share, and the following agents merge their own information into the chain and then forward the new message to the next agent. The chain ends when the final agent finds the message useless for others, or there are no more potential communication objects.

When designing the communication protocol, we mainly consider the following questions: (1) Whether to communicate? (2) Whom to communicate with? (3) What to communicate? (4) How to utilize the messages?

\begin{figure*}[t]
% \vspace{-0.5cm}
\centering
\includegraphics[width=0.9\textwidth]{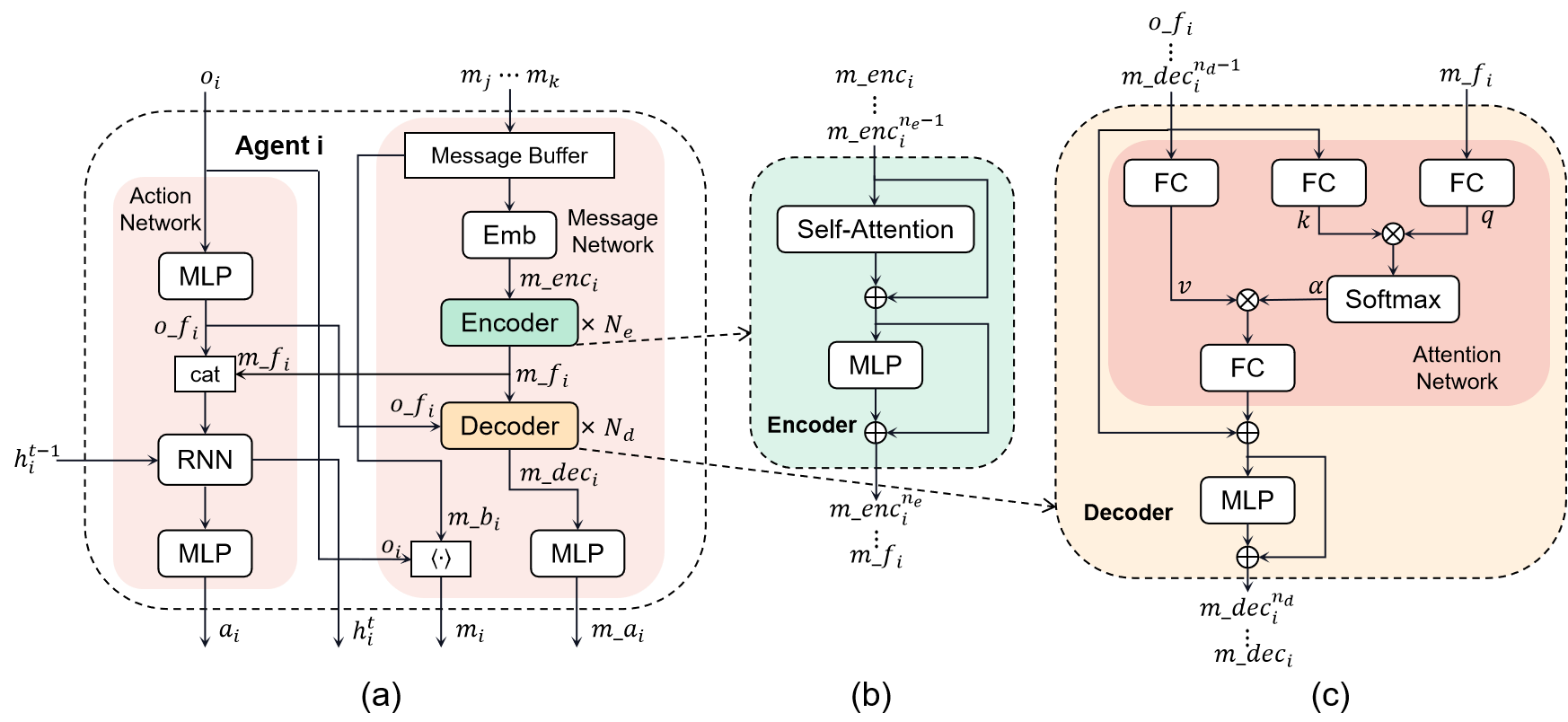} % Reduce the figure size so that it is slightly narrower than the column.
\caption{Network structure of TEM. (a) Actor network of agent \textit{i}, including an action network and a message network. Emb denotes the embedding network. (b) Encoder module. (c) Decoder module, where $m\_dec_i^{0}=o\_f_{i}$.}
\label{fig2}
\vspace{-0.5cm}
\end{figure*}

\textbf{(1) Whether to communicate?}
As shown in Fig \ref{fig1}, in every step of execution and training, the first stage is communication. When the communication stage is done, the actor networks for each agent will make the action decisions by the observations and messages. In the communication stage, each agent has the chance to decide whether to start a new chain and send a message. And the agents who receive messages can decide whether to continue forwarding the messages. Multiple message chains are allowed and the information from different chains is merged if a shared node exists.

\textbf{(2) Whom to communicate with?}
For partial-observation (PO) problems, communication with all the agents is not reasonable and effective. The direct communication with the agent outside the observation range may not bring helpful information as the sender does not even know the receiver's state. Therefore, we do not model all the other agents to decide whether to communicate with them like in some previous works \cite{ding_learning_2020,yuan_multi-agent_2022}. Instead, when the agent \textit{i} chooses communication objects, we only consider the agents in the observation range $\mathcal{O}_i$ , and in our experiments, $\mathcal{O}_i$ includes the nearest several agents of the agent \textit{i}. By training the message module, the agent can predict the impact of the message on other agents, and is more likely to choose the one with the highest impact to communicate.

We combine the two decisions (1) and (2) into one communication action. The communication actions of agent \textit{i} $m\_a_i$ include not sending at all $m\_a_i=0$, and sending to one agent \textit{j} in the observation range $m\_a_i=j,~ (j\in \mathcal{O}_i)$. Namely:
\begin{equation}
\label{eq4}
\begin{aligned}
P(m\_a_i=0)+\sum_{j\in \mathcal{O}_i}\left(P(m\_a_i=j)\right)=1.
\end{aligned}
\end{equation}
% \begin{equation}
% \label{eq4}
% \begin{aligned}
% 1=&\sum_{j\in \mathcal{O}_i}\left(P(m\_a_i=j)\right)+\\
% &P(m\_a_i=0).
% \end{aligned}
% \end{equation}

This way, the agent can decide when and whom to 
% \vspace{-0.3cm}
\begin{algorithm}[htbp]
% \tiny
    \caption{Execution workflow of agent $i$ in one time step}
    \label{alg:algorithm}
    % \textbf{Input}: Your algorithm's input\\
    % \textbf{Parameter}: Optional list of parameters\\
    % \textbf{Output}: Your algorithm's output
    \begin{algorithmic}[1] %[1] enables line numbers
        \STATE $stop\_flag_i=0$, $m\_re_i=$Null, $sender\_list_i=$Null
        \FOR{$com\_step_i = 0 \to max\_com\_steps$}
        \STATE Merge the received message $m\_re_i$ to the buffer $m\_b_i$, and add the senders of $m\_re_i$ to $sender\_list_i$
        \IF {$stop\_flag_i==0$}
        \STATE Get the target $m\_a_i$ by $o_i$ and $m\_b_i$
        \IF {$m\_a_i$ in $sender\_list_i$}
        \STATE $stop\_flag_i=1$
        \ENDIF
        \IF {$m\_a_i\ne0$ and $stop\_flag_i\ne1$}
        \STATE Send the message to the agent $m\_a_i$ as $m\_re_{m\_a_i}$
        \ENDIF
        \ENDIF
        \ENDFOR
        \STATE Compute the action $a_i$ by $o_i$ and $m\_b_i$
        \STATE Interact with the environment to get the new $o_i$
    \end{algorithmic}
\end{algorithm}
% \vspace{-0.3cm}
communicate by one action, simplifying the modeling and learning.

\textbf{(3) What to communicate?}
To keep the information from the head nodes in the chain, and merge the information from different chains, every agent maintains a message buffer to store the messages. In practice, the message buffer is implemented as a queue, with a fixed storage length, but can flexibly push in and pop out elements as the communication goes (First Input First Output, FIFO). We use $m\_b_i$ to denote all the messages inside the agent \textit{i} 's message buffer. When sending the new message, the agent \textit{i} merges its own observation into the chain, then the message chain expands to $\langle m\_b_i, o_i \rangle$. Here, the operation $\langle \cdot \rangle$ denotes pushing into the queue to concatenate the messages. The buffer is clear when every step starts.

\textbf{(4) How to utilize the messages?}
Instead of some previous works \cite{zhang_efficient_2019,yuan_multi-agent_2022}, we do not think the messages directly influence the value estimation of other agents is the natural way of communication. The information exchange should be separated from the information utilization. And the final effects of the messages should be determined by the receiver instead of the sender. Thus, in our model, messages are taken as a counterpart of the observation, serving as part of the inputs of the actor network.

\subsection{Network Design}
The schematics of the network design in our model are shown as Fig \ref{fig2}. Each agent has an actor network to observe the environment and communicate with other agents. The actor network of the agent \textit{i} will output the action to interact with the environment $a_i$, the action to communicate $m\_a_i$ (whether to communicate and whom to communicate with), and the corresponding message to be sent $m_i$. To better utilize the history information and get a smoother action sequence, an RNN is employed in the actor network. Thus the agent \textit{i} also keeps a hidden state $h_{i}^{t}$, and updates it every time step.

The actor network consists of two sub-networks, the action network and the message network. The action network mainly concentrates on the task itself and tries to get better rewards by outputting reasonable actions. The message network concentrates on the communication to share information with other agents instead. The two sub-networks exchange the representation feature of the observations $o\_f_{i}$ and that of the messages $m\_f_{i}$ to merge the information. 

In the action network, $o\_f_{i}$ is learned by a multi-layer perceptron (MLP), and then the action network concatenates $o\_f_{i}$ and $m\_f_{i}$ to input into the RNN together with the hidden state from the last time step $h_{i}^{t-1}$. Another MLP, in the end, processes the output of the RNN to get the action $a_i$. 

On the other hand, the messages from other agents like $m_j \cdots m_k$ are stored in the message buffer, like the email inbox. The embedding layer (we implement it as a full connected (FC) layer by practice) converts the messages to fit the input dimensions of the encoders. $N_e$ sequential encoder modules and $N_d$ sequential decoder modules are followed by the embedding layer. The output of encoder modules $m\_f_{i}$ serves as the representation of all the messages in the buffer, with the key information emphasized by the attention mechanism. The decoder modules further combine the information from both of $m\_f_{i}$ and  $o\_f_{i}$ to get the output $m\_dec_i$. Finally, one MLP produces the communication decision $m\_a_i$.

For each encoder module, it takes in $m\_enc_i^{n_e-1}$ from the embedding layer or the last encoder, then generates $m\_enc_i^{n_e}$ as the input for the next layer. The transformer in the module can model the sequential information and is flexible to fit message chains with different lengths. Also, the attention mechanism will help the agent to pick out the key information from the chain. $n_e$ implies the position of the layer in the encoder sequence. To prevent gradient vanishing, the encoder module employs the residual connections to link the self-attention mechanism and the MLP \cite{wen_multi-agent_2022}. The structure of the self-attention mechanism is the same as the attention network in the decoder while $k, q$ and $v$ are generated from the same input $m\_enc_i^{n_e-1}$. 

In the decoder module, the first $m\_dec_i^{0}$ is the representation of the observations $o\_f_{i}$. In the attention network, full connected layers generates key $k$ and query $q$ by $m\_dec_i^{n_d-1}$ and $m\_f_i$, respectively. Also, the third FC layer generates value $v$ from $m\_dec_i^{n_d-1}$. $k$ and $q$ are used for calculating the weights $\alpha$ of the value $v$ as Equation \ref{eq1}.
\begin{equation}
\label{eq1}
% \alpha=\frac{1}{\lambda}exp(\frac{\bm{q}\bm{k}^T}{\sqrt{d_k}}),
\alpha=Softmax(exp(\frac{\bm{q}\bm{k}^T}{\sqrt{d_k}})).
\end{equation}
% where $\lambda=\sum_$
In fact, the weight $\alpha$ learns the correlations between the $m\_dec_i^{n_d-1}$ and $m\_f_i$. By multiplying $v$ and $\alpha$, then we get the weighted representation of $m\_dec_i^{n_d-1}$ from the ending FC layer. With a similar structure of the residual connections and MLP, we get $m\_dec_i^{n_d}$ as the input for the next layer. 
% CDTE
\subsection{Loss Function Design}
The communication among the agents in a collaborative task aims to share the key information that one believes is useful for some specific agents. So the learning of the message network is driven by the impact of the message to be sent.  

As the communication will not change either the action of other agents or the loss of the action network, an independent loss to model the influence of the messages on other agents' actions is needed. We denote the communication loss as $\mathcal{L}^{(i)}_m(\theta)$, where $\theta$ is the parameters of the actor network. The action of an agent \textit{j} is sampled from the categorical distribution $P(a_j|o_j,m\_b_j)$ learned by the action network. Then, when considering the new message from the agent \textit{i} $m_i$, we can estimate the distribution $P(a_j|o_j,\langle m\_b_j, m_i \rangle)$ as the consequence of the communication. Kullback-Leibler (KL) divergence is widely used to measure the discrepancy between these two conditional probability distributions. Thus, the causal effect $\Gamma_j^{(i)}$ of the message from agent \textit{i} on agent \textit{j} can be defined as:
\begin{equation}
\label{eq2}
\Gamma_j^{(i)}=D_{KL}\left(P(a_j|o_j,\langle m\_b_j, m_i \rangle )||P(a_j|o_j,m\_b_j)\right).
\end{equation}
By considering all the possible agents to send the message to in the observation range, we can get the expectation of the causal effect of the message $\mathbb{E}\Gamma^{(i)}(\theta)$ by Equation \ref{eq3}:
\begin{equation}
\label{eq3}
\begin{aligned}
\mathbb{E}\Gamma^{(i)}(\theta)
&=\sum_{j\in \mathcal{O}_i}\left(P_{\theta}(m\_a_i=j|o_i,m\_b_i)\Gamma_j^{(i)}\right)
% \\
% &=\sum_{j\in \mathcal{O}_i}\left(P_{\theta}(m\_a_i=j|o_i,m\_b_i)\\
% &D_{KL}(P(a_j|o_j,\langle m\_b_j, m_i \rangle)||P(a_j|o_j,m\_b_j))\right)
,
\end{aligned}
\end{equation}
where $\mathcal{O}_i$ denotes the observation range of the agent \textit{i}. The communication decision of agent \textit{i} is sampled form the categorical distribution $P_{\theta}(m\_a_i|o_i,m\_b_i)$ learned by the message network. $P_{\theta}$ denotes that the gradient of this item should be propagated when training. 

The expectation $\mathbb{E}\Gamma^{(i)}(\theta)$ represents the overall effect the message $m_i$ can bring to the whole system, which we should maximize in the loss function. However, communication should also be sparse and efficient. If we do not control the communication times by the external guidance, the agents will tend to send as many messages as possible to get higher $\mathbb{E}\Gamma^{(i)}(\theta)$. Therefore, we also designed another item for communication loss to reduce the communication overhead. When the agent \textit{i} chooses not to send the message to any agents in the observation range for most of the times, the probability $P_{\theta}(m\_a_i=0|o_i,m\_b_i)$
should be relatively high. So we need to maximize this probability at the same time.

So far, we can get the final communication loss $\mathcal{L}^{(i)}_m(\theta)$ by the following equation and maximize it when training.
\begin{equation}
\label{eq6}
\mathcal{L}^{(i)}_m(\theta)=\mathbb{E}\Gamma^{(i)}(\theta)+\delta P_{\theta}(m\_a_i=0|o_i,m\_b_i),
\end{equation}
where $\delta$ is the weight of the communication reduction.

The loss of the action network $\mathcal{L}^{(i)}_a(\theta)$ is defined followed by Equation \ref{La} in MAPPO as:
\begin{equation}
\label{eq4}
\begin{aligned}
\mathcal{L}^{(i)}_a(\theta)=&min[r_{\theta}^{(i)}A^{(i)}_{\pi_{\theta_{old}}}, clip(r_{\theta}^{(i)},1-\epsilon,1+\epsilon)A^{(i)}_{\pi_{\theta_{old}}}],
\end{aligned}
\end{equation}
where $r_{\theta}^{(i)}=\frac {\pi_\theta(a^{(i)}|o^{(i)})}{\pi_{\theta_{old}}(a^{(i)}|o^{(i)})}$,  $A^{(i)}_{\pi_{\theta_{old}}}$ is the advantage function.

What's more, to encourage more exploration when training, we adopt an entropy loss $\mathcal{L}^{(i)}_e(\theta)$ as \cite{yu_surprising_2021}:
\begin{equation}
\label{eq6}
\mathcal{L}^{(i)}_e(\theta)=S(\pi_{\theta}(o_i)).
\end{equation}
We can get the overall loss function for the actor network when training:
\begin{equation}
\label{eq7}
\begin{aligned}
\mathcal{L}(\theta)=\sum_{i=1}^n \left( \mathcal{L}^{(i)}_a(\theta) + \lambda_m \mathcal{L}^{(i)}_m(\theta) + \lambda_e \mathcal{L}^{(i)}_e(\theta)\right),
\end{aligned}
\end{equation}
where $n$ is the number of the agents, and $\lambda_m, \lambda_e$ are the coefficients to weight the corresponding losses.

The loss of critic network keeps the same as MAPPO.
% \begin{equation}
% \label{eq8}
% \begin{aligned}
% \mathcal{L}(\phi)=& \sum_{i=1}^n (max[(V_\phi(s^{(i)})-R)^2,\\
% &(clip(V_\phi(s^{(i)}),V_{\phi_{old}}(s^{(i)})-\epsilon,V_{\phi_{old}}(s^{(i)})+\epsilon)-R)^2]),
% \end{aligned}
% \end{equation}

% \begin{equation}
% \label{eq8}
% \begin{aligned}
% \mathcal{L}(\phi)=& \frac{1}{B_n}\Sigma_{i=1}^{B}\Sigma_{k=1}^{n}(max[(V_\phi(s_i^{(k)})-R_i)^2,\\
% &(clip(V_\phi(s_i^{(k)}),V_{\phi_{old}}(s_i^{(k)})-\epsilon,V_{\phi_{old}}(s_i^{(k)})+\epsilon)-R_i)^2])
% \end{aligned}
% \end{equation}
% where $R$ is the discounted accumulated reword. 
% and the $B$ refers to the batch size.

\begin{figure}[]
% \vspace{-0.5cm}
\centering
\includegraphics[width=0.95\columnwidth]{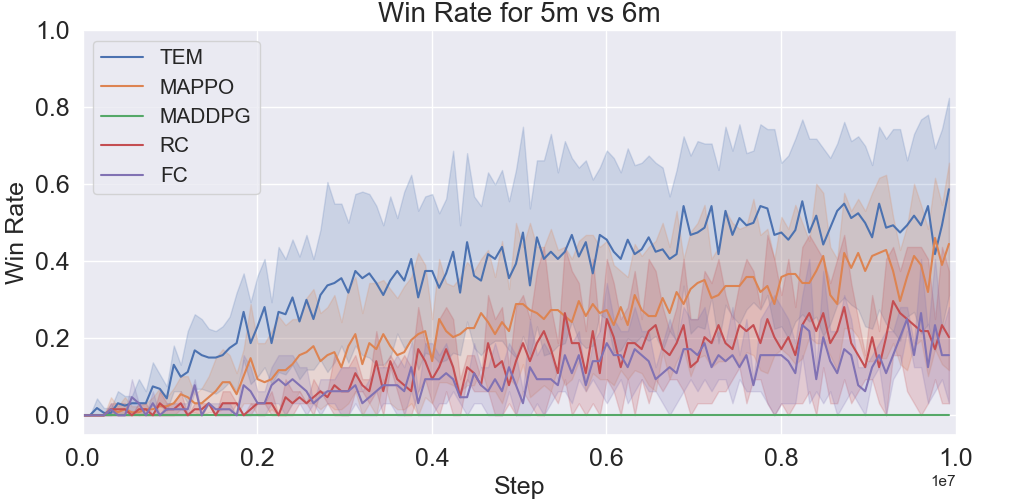} % Reduce the figure size so that it is slightly narrower than the column.
\caption{Test win rate for the SMAC map 5m vs. 6m, the shaded regions represent the 95\% confidence intervals over 5 seeds. FC: Full Communication, RC: Randomly-stop Communication.}
\label{figsmac}
\vspace{-0.4cm}
\end{figure}

\section{Experiments}
We evaluate the performance of TEM on three widely-used partially-observed multi-agent cooperative tasks: the Starcraft Multi-Agent Challenge (SMAC), Predator Prey (PP) and Cooperative Navigation (CN). We compare the training process of TEM with the baselines and analyze the performance. We test the scalability of TEM to scenarios with different numbers of agents and targets when zero-shot transferring. 
% \subsection{Environments and Baselines}
\vspace{-0.5cm}
\subsection{Starcraft Multi-Agent Challenge}
\vspace{-0.1cm}
StarCraft II is a real-time strategy game serving as a benchmark in the MARL community. In the Starcraft Multi-Agent Challenge (SMAC) task, N units controlled by the learned algorithm try to kill all the M enemies,
% .There are usually more enemies than agents
% , or the enemies are more powerful types of units, 
% so defeating all the enemies is challenging
demanding proper cooperation strategies and micro-control of movement and attack. We choose the hard map 5m vs. 6m to evaluate TEM. TEM controls 5 Marines to fight with 6 enemy Marines.

The baselines include MAPPO, MADDPG, Full Communication (FC) and Randomly-stop Communication (RC). MAPPO is the CTDE backbone we are using in the following experiments, which is proven to have state-of-the-art performance on several MARL cooperative benchmarks \cite{yu_surprising_2021}. MADDPG is another classic CTDE approach for multi-agent cooperation tasks \cite{lowe_multi-agent_2020}. FC and RC are two special cases of TEM. We keep the communication protocol the same, but disable the decoder in the message module, instead, the agents choose the communication targets by pre-defined rules. In FC, the agent will keep randomly choosing someone to communicate with, to extend the message chain until no one is available. In RC, the agent will randomly stop the message chain by a probability $p$, or keep forwarding to a random one.

We run the experiments over five seeds. For each seed, we compute the win rate over 32 test games after each training iteration as shown in Fig. \ref{figsmac}. TEM gets the highest win rate over the baselines. FC and RC perform worse than MAPPO benchmark. One possible reason is that targeted communication by TEM could improve cooperation while random communication by FC and RC may bring redundant information for decision-making. 
The win rate of MADDPG remains zero, showing that it is hard to defeat an army with more units.
% The win rate of baseline MADDPG remains zero, showing that it is hard to defeat an army with more units and MADDPG fails to learn such a strategy. 

\subsection{Predator Prey}

\begin{figure}[]
% \vspace{-0.5cm}
\centering
\includegraphics[width=0.95\columnwidth]{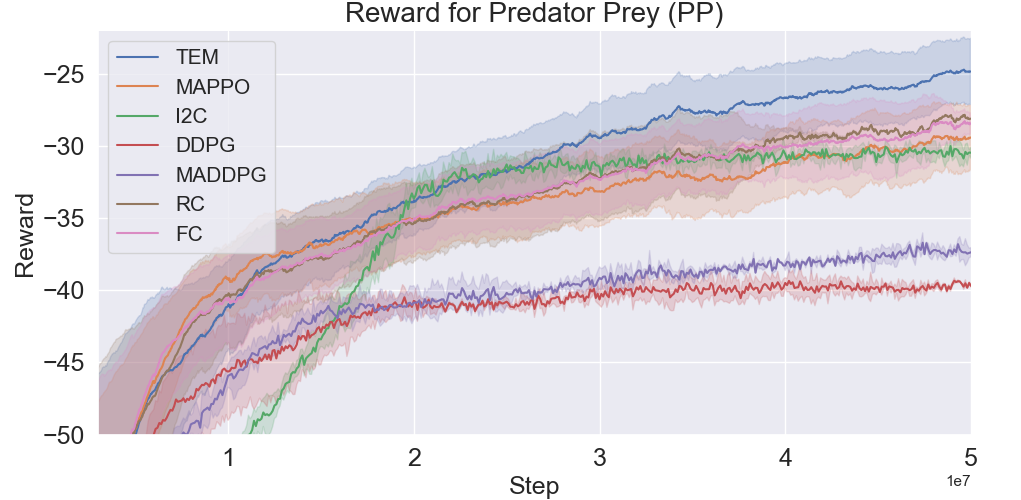} % Reduce the figure size so that it is slightly narrower than the column.
\caption{Learning curves for Predator Prey (PP)
% , the shaded regions represent the 95\% confidence intervals over 5 seeds
.}
\label{fig3}
\vspace{-0.2cm}
\end{figure}
In the Predator Prey (PP) task, N predators try to chase, surround and finally capture M preys, as shown in Fig \ref{fig0}. The predators are the agents to be trained and the preys flee in the opposite direction of the closest predator at a faster speed following pre-defined rules. So the predators have to be grouped automatically and cooperate to surround each prey, and it is impossible for one predator to capture a prey itself. In practice, we set N as 7 and M as 3, denoted as 7-3 scenario. 

Different from the PP task in some previous works, here, the agents can only partially observe the teammates and targets. The rewards are the sum of the agents' negative distances to their closest preys or landmarks. In addition, the agents are penalized for collisions with other agents. 

The baselines include MAPPO, I2C, MADDPG, DDPG, FC and RC. I2C proposes an individual communication mechanism \cite{ding_learning_2020}. DDPG is a classic deep reinforcement learning algorithm for continuous control \cite{lillicrap_continuous_2019}. We apply DDPG independently to each agent as a baseline without considering cooperation.

As shown in Fig \ref{fig3}, while other baselines gradually converge at the last episodes, TEM keeps raising the rewards and improves the final reward by 17.2\% compared with MAPPO. 

\begin{figure}[bp]
\vspace{-0.2cm}
\centering
\includegraphics[width=0.95\columnwidth]{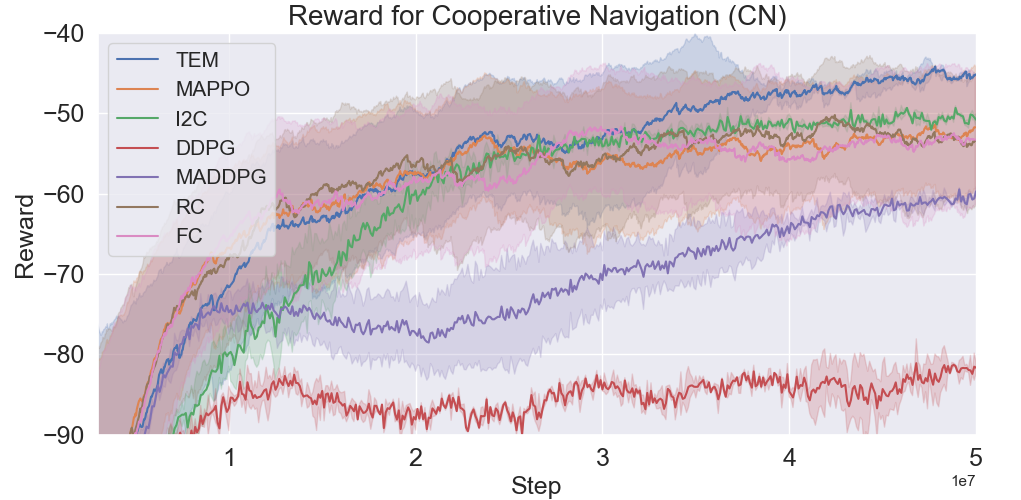} % Reduce the figure size so that it is slightly narrower than the column.
\caption{Learning curves for Cooperative Navigation (CN)
% , the shaded regions represent the 95\% confidence intervals over 5 seeds
.}
\label{fig4}
% \vspace{-0.5cm}
\end{figure}

% \subsection{Benefits from Communication}
\subsection{Cooperative Navigation}

\begin{figure}[]
% \vspace{-0.7cm}
\centering
\includegraphics[width=0.9\columnwidth]{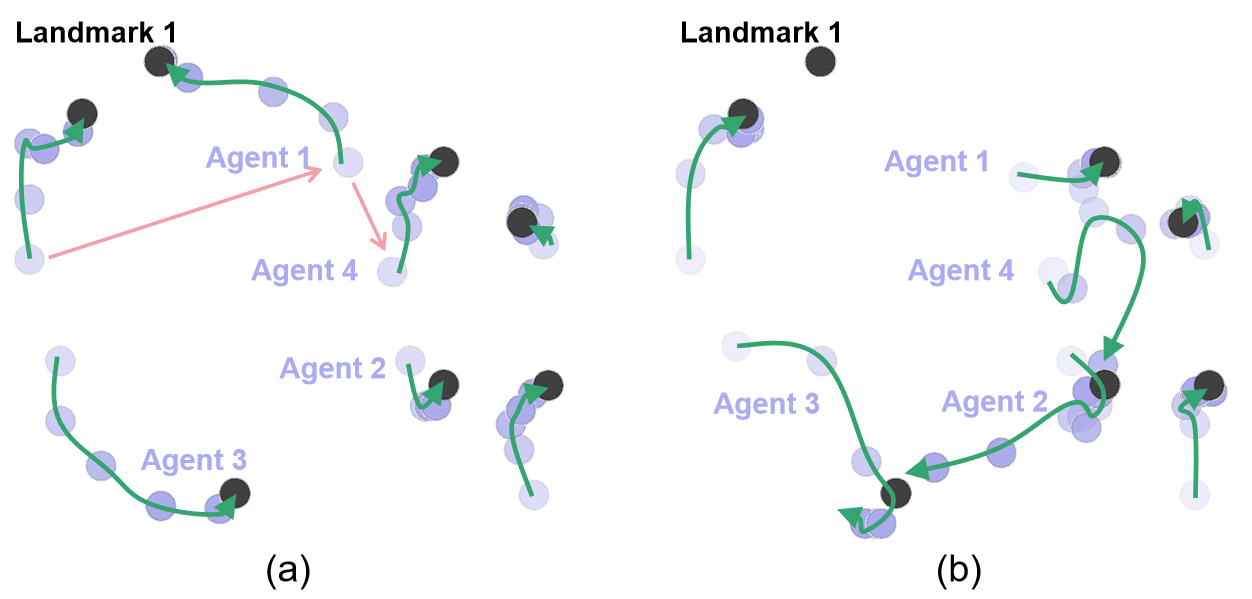} % Reduce the figure size so that it is slightly narrower than the column.
\caption{
% Comparison between (a) TEM and (b) MAPPO on the same environment of CN. Five timesteps are illustrated. The darker agents demote the later timestep. The green lines are the trajectories of the agents (purple) to occupy the landmarks (black). The pink lines are message chains.
Comparison between (a) TEM and (b) MAPPO on CN. Five timesteps are illustrated. Green: trajectories of the agents (purple) to occupy the landmarks (black). Pink: message chains.
}
\label{fig6}
\vspace{-0.3cm}
\end{figure}

In the Cooperative Navigation (CN) task, N agents try to occupy N stationary landmarks separately, as shown in Fig \ref{fig6}. The positions of landmarks and agents are randomly initialized. The best strategy is that each agent has a different target from the beginning through communication instead of rescheduling when collisions happen because of choosing the same target. In practice, we set N as 7, denoted as 7-7 scenario. The baselines and reward settings are the same as PP.

% \begin{figure}[!b]
% \vspace{-0.5cm}
% \centering
% \includegraphics[width=1\columnwidth]{example.png} % Reduce the figure size so that it is slightly narrower than the column.
% \caption{Illustration of TEM on (a)\&(b): PP and (c)\&(d): CN. The red and green dotted lines denote the trajectories of the preys (black) and the agents (purple), respectively. The blue arrows are the main directions of the agents to surround the preys. The pink lines are the message chains. }
% \label{fig5}
% \vspace{-0.5cm}
% \end{figure}
We compare TEM with the baselines on the training performance Fig \ref{fig4}. We can see that TEM converges to the highest reward than all the baselines. FC and RC are only slightly better than MAPPO, suggesting that the communication actions $m\_a$ learned by TEM are targeted, and the message chain brings helpful information to the ones that really need it. 

% We illustrate the details of how TEM works on PP and CN in Fig \ref{fig5}. Figure (a) is the first step of PP, where two message chains are formed. The agents of Chain 1 start to move down and right as they find all the observed preys are in these directions by sharing the information. In a following step (b), the agents of Chain 3 keep moving right to drive the preys to the bottom right corner. The agents of Chain 4 reach a similar consensus to surround all the preys at the corner. From the red dotted trajectories of the preys, we can see that the three preys are close and trapped in the encirclement in the end.

% In Fig \ref{fig5} (c) and (d), the two message chains are coordinating the agents on the left and right, respectively. It is noted that Agent 1 finds there is no landmark left on the right by Chain 6 in (c), so in a following step (d), Agent 1 moves left and joins Chain 7 and finally occupies a bottom left landmark. What's more, Agent 2 and Agent 3 learn to leave the nearest landmarks to the agents far away but to occupy other landmarks, so that the whole team can occupy all the landmarks as soon as possible.

We compare the illustrations on CN between TEM and MAPPO in Fig \ref{fig6}. In (a), the TEM agents Agent 1 and Agent 4 notice Landmark 1 by communication (pink message chain). Thus each agent moves straight forward to the corresponding landmarks.  While in (b), the MAPPO agents miss Landmark 1, so for Agent 4, there will be nowhere to go. Agent 4 first tries to scramble with Agent 1 but fails, then turns to Agent 2. Agent 2 is forced to leave to avoid collision and turns to Agent 3. We can see that communication brought by TEM can improve cooperation and reduce internal strife.

\subsection{Scalability of TEM}
We further examine the scalability of TEM on PP in Table \ref{tab1}.   We take average episode rewards (R), successful capture times (S), collision times (C) as the metrics. For R and S, the performance is better when the values are higher, while for C, the performance is better when the values are lower. Note that the existing selective MARL communication approaches are not scalable due to the modeling of each agent, so the baselines are the transferred MAPPO and the specifically trained MAPPO. We directly transfer the learned model from the 7-3 scenario to 9-3 and 3-1 scenarios without further training. For 9-3 scenario, two new agents are included so the task will be easier. But more agents also increase the risks of collision, so the cooperation mode could be different and the agents need to communicate to suit the new scenario. For TEM, the average episode rewards rise from -40.5 to -17.9, and the gain is 55.8\%, while for MAPPO, the gain of rewards is 52.3\%. TEM does not only perform better after transferring, but also gains more.
\begin{table}[!htbp]
% \vspace{-0.7cm}
	\centering\tiny
 % \tiny
	\begin{threeparttable}
		\begin{tabular}{cccccc}
			\toprule
			 &&\textbf{TEM} (7-3)& MAPPO (7-3)& MAPPO (learned) & \textbf{TEM} (finetuned)\\
			\midrule
			\multirow{3}{*}{7-3} &R & -40.5$\pm$4.7 & -44.9$\pm$4.3 & - & -    \\
			 &S & 61.6$\pm$18.3 & 49.0$\pm$16.5 & -  & -  \\
			 &C & 1.4$\pm$0.6 & 12.6$\pm$2.6 & -  & -  \\
			\midrule\midrule
			\multirow{3}{*}{3-1} &R & -10.6$\pm$4.7 & -12.7$\pm$5.9 &  -7.52$\pm$2.6 &-7.0$\pm$2.8\\
			 &S & 18.7$\pm$14.0 & 13.5$\pm$12.7  &  36.5$\pm$13.2  &31.7$\pm$13.4\\
			 &C & 1.2$\pm$1.2 & 3.6$\pm$2.7 &  1.8$\pm$1.8 &0.8 $\pm$0.6\\
			 \midrule
			\multirow{3}{*}{9-3} &R & -17.9$\pm$5.0 & -21.3$\pm$7.5 & -17.1$\pm$ 8.2 &-14.0  $\pm$2.6\\
			 &S & 107.2$\pm$23.6 & 69.3$\pm$32.0 &  109.5$\pm$18.7    &127.9$\pm$5.9\\
			 &C & 16.2$\pm$1.8 & 30.6$\pm$6.7 &  7.2$\pm$1.9   &5.4$\pm$1.6\\

\bottomrule
		\end{tabular}
	\end{threeparttable}
\caption{Scalability of TEM on PP. R: average episode rewards, S: successful capture times, C: collision times. TEM (7-3) and MAPPO (7-3) are trained on the scenario 7-3: 7 agents to capture 3 preys, and tested on ten random environments on 7-3, 3-1, 9-3 scenarios. MAPPO (learned) is specifically trained from scratch on the corresponding test environments. TEM (finetuned) is the TEM model trained on 7-3 and tuned on the corresponding test environments.}
\label{tab1}
\vspace{-0.4cm}
\end{table}	
For 3-1 scenario, both the numbers of the agents and preys change. The results show that TEM still keeps a better performance on all the metrics. Moreover, it shows that after TEM learns how to communicate in a complex scenario, it can successfully transfer to simple ones.

We also train MAPPO from scratch specifically on 9-3 and 3-1 (denoted as MAPPO (learned)), and the performance of transferred TEM (trained on 7-3) is close to MAPPO (learned) on 9-3 without training. But the transferred TEM works worse on 3-1, and we suggest that cooperation by communication may not play an essential role in such a simple environment. We further finetune TEM (7-3) on the new scenarios and the finetuned models even outperform the specially learned MAPPO.

Similar experiments are conducted on CN as shown in Table \ref{tab2}. TEM keeps the scalability when transferred from 7-7 scenario to 6-6 and 9-9, and outperforms the transferred MAPPO. Surprisingly, the transferred TEM even outperforms the MAPPO trained from scratch (denoted as MAPPO (learned)) on most metrics. It suggests that CN requires more communication to coordinate the agents to explore all the landmarks. The results also show that the communication pattern learned from 7-7 still works well in other scenarios. Similarly, the finetuned TEM gets even better performance.

\begin{table}[!htbp]
\vspace{-0.2cm}
	\centering\tiny
	\begin{threeparttable}
		\begin{tabular}{cccccc}
			\toprule
			 &&\textbf{TEM} (7-7)& MAPPO (7-7)& MAPPO (learned)& \textbf{TEM} (finetuned) \\
			\midrule
			\multirow{3}{*}{7-7} &R & -38.8$\pm$15.1 & -46.6$\pm$14.8 & - & -   \\
			 &S & 35.8$\pm$ 6.2& 23.3$\pm$9.0 & - & -   \\
			 &C & 2.8$\pm$0.3 & 4.2$\pm$0.2 & -  & -  \\
			\midrule\midrule
			\multirow{3}{*}{6-6} &R & -39.8$\pm$5.3 & -45.0$\pm$8.0 & -43.6$\pm$10.1 &-36.7  $\pm$5.2\\
			 &S & 35.3$\pm$6.2 & 20.0$\pm$5.9  &  19.3$\pm$6.8  &36.1$\pm$5.9\\
			 &C & 7.2$\pm$0.4 & 8.4$\pm$0.4 & 4.8$\pm$0.4 &4.8  $\pm$0.2\\
			 \midrule
			\multirow{3}{*}{9-9} &R & -45.8$\pm$23.9 & -57.5$\pm$25.3 &  -50.8$\pm$12.9 &-41.1 $\pm$11.4\\
			 &S & 39.4$\pm$4.9 & 26.6$\pm$8.7 & 29.0$\pm$6.6     &38.9$\pm$5.2\\
			 &C & 9.0$\pm$0.3 & 28.8$\pm$0.4 &  10.8$\pm$0.2   &8.0$\pm$0.2\\

\bottomrule
		\end{tabular}
	\end{threeparttable}
\caption{Scalability of TEM on CN. 
TEM (7-7) and MAPPO (7-7) are trained on the scenario 7-7: 7 agents to occupy 7 landmarks, and tested on ten random environments on 7-3, 6-6, 9-9 scenarios.
}
\label{tab2}
\vspace{-0.5cm}
\end{table}			
\section{Conclusions}
To tackle the scalability problem of MARL communication, this paper proposes a novel framework Transformer-based Email Mechanism (TEM). The agents adopt local communication to send and forward messages like emails to form message chains, which set up bridges among partial-observation ranges. We introduce Transformer to encode and decode the message chain to choose the next receiver selectively. Empirical results in diverse multi-agent cooperative tasks show that our method outperforms
the baselines. Furthermore, we can directly apply TEM to a new environment with a different number of agents without retraining. Better performance than the baselines when zero-shot transferring shows the scalability of TEM. Based on TEM, communication for hundreds of agents and further tailored message generation can be developed, which may be an important step for MARL applications to real-world tasks.

\section*{Acknowledgments}
This work was supported by the National Key Research \& Development Program of China (No. 2021YFC1809003).
\bibliographystyle{named}
% \bibliography{ijcai22}
% \begin{quote}
% \begin{small}
\bibliography{references.bib}
% \end{small}
% \end{quote}

\end{document}